\documentstyle[aps,preprint]{revtex}
\begin{document}
\bibliographystyle{unsrt} 

\title{On the lowest energy excitations  of one-dimensional strongly correlated electrons} 

\author{Konstantin Kladko}
\address{Max-Planck-Institute for Physics of Complex Systems,\\ 
Noethnitzer Str. 38, D-01187 Dresden, Germany\\
and\\
Theoretical Division, MS-B258,  Los Alamos National Laboratory, \\ 
Los Alamos, 87545, New Mexico, USA \\
e-mail:kladko@lanl.gov \\
homepage: http://cnls.lanl.gov/$\sim$kladko/}

\maketitle
\begin{abstract}
It is proven that the lowest excitations $E_{low}(k)$ of one-dimensional half-integer spin  
generalized Heisenberg models and half-filled  extended Hubbard models   
are  $\pi$-periodic functions. For Hubbard models at fractional fillings $E_{low}{(k+ 2 k_f)} = E_{low}(k)$,  
where $2 k_f= \pi n$, and $n$ is the number of electrons per unit cell. 
Moreover, if one of the ground states of the system is magnetic in the thermodynamic limit, then
$E_{low}(k) = 0$ for any $k$, so the spectrum is gapless at any wave vector. The last statement is true for 
any integer or half-integer value of the spin. 
\end{abstract}
{
Low-energy excitation spectra of correlated systems have long been in the center of the theoretical
condensed matter research (see \cite{fulde} and references therein). In 1932 Hans Bethe proposed 
his Ansatz method \cite{bethe} for correlated wave functions, that was then used to
solve many one-dimensional quantum and also two-dimensional classical models (see \cite{korepin}, \cite{mattis},
\cite{korepin2}).
Using the Bethe Ansatz (BA) method des Cloizeaux and Pearson \cite{pearson} derived  the energy of  
low lying $S=1$ (triplet) states of 
the spin-1/2 antiferromagnetic Heisenberg model (AFHM). The spectrum of the excitations  was found to be 
 $E_{t}\left( k \right)= 1/2 J \pi \mid \sin k \mid $. 
$E_t\left( k \right)$ is a $\pi$-periodic function
of $k$. 
des Cloizeaux and Pearson assumed the excitations $E_t\left( k\right) $ to be the  elementary excitations of the system,
therefore assigning to an elementary excitation spin 1. Later investigations \cite{johnson} 
have shown that the spectrum of 
$\cite{pearson}$ was incomplete. The full spectrum of the  $S=1$ magnon excitations is characterized by two continuous quantum
numbers. Fixing the wave vector $k$ one still has a continuous one-parametric family of excitations.
The des Cloizeaux and Pearson spectrum is  only a lower bound of these excitations.  As a result $S=1$ states can  
not be considered as elementary indivisible particles having only one wave number $k$. They are rather two-particle
compositions. The problem was finally clarified 
by Faddeev and Takhtajan in Ref. \cite{faddeev}. It was  shown that all excitations of the 
spin-1/2 AFHM are superpositions of spin-1/2 elementary excitations (called kinks
or spinons).
For periodic rings having  odd/even number of sites only odd/even number of kinks are possible in the
system. Kinks do not form bound states. Their interaction manifests itself only through a scattering amplitude. 
Energies and momenta of many-kink states add, as for  independent particles.
The dispersion relation of a kink is $1/2 J \pi \sin k$, with  $0\le k \le \pi$. 
The Brillouin 
zone is therefore only half of the original one. The $S=1$ states of the model are  pairs of kinks.
The  lowest excitations are found by minimizing $E\left( k_1\right)  + E\left( k_2\right) $ for $k_1+k_2=k$. The result is
$k_1=k$ and $k_2=0$ (or vice versa), so the lowest excitations probe
the one-spinon excitations spectrum.      
For the ferromagnetic case the one magnon spectrum of the one-dimensional spin-1/2 ferromagnetic Heisenberg model (FHM) 
has the form $2 J ( 1 - \cos k)$, possessing a gap at nonzero $k$ and not being  $\pi$-periodic.
Careful analysis shows, that one-magnon states are not the lowest energy states of the system. Magnons
can form bound states and the interaction is so strong,, that the two-magnon states have energy less
than the energy of one magnon. It was shown, using Bethe Ansatz  \cite{faddeev2}, that the elementary
excitations of the system (excitations having one quantum number) are magnons and strings. 
Strings are complexes of $2M+1$ spins, having a  dispersion relation $J \frac{2}{2 M +1} (1-\cos k )$.  
The lowest bound for the spectrum of excitations is formed, when $M$ goes to infinity. Therefore the 
spectrum is gapless for all wavevectors $k$. As will be shown below, this property characterizes {\it all}
 Heisenberg-like one-dimensional models (integer-spin or half-integer-spin), which are magnetic in the thermodynamic limit.  
The excitations spectrum of the half-filled one-dimensional Hubbard model was derived in \cite{ovchinnikov}.
The lowest-lying triplet  states of the model are given by equation (16) of \cite{ovchinnikov}.
The function $E_{t} \left( k \right) $ is $\pi$-periodic and reaches its maximum at $q=\pi/2$. For a quarter filling 
the spectrum of the lowest energy excitations is periodic with a period of $\pi/2 = 2 k_f$, where $k_f$ is the
Fermi wave vector of the noninteracting system. For the attractive Hubbard model the excitations spectrum
is calculated in \cite{woynarovich}.  The elementary excitations of the model at half-filling 
are  
bound pairs and "free" electrons (instead of spinons and holons in the repulsive case). The energy-momentum dispersion relation
of  excitations is the same, as in the repulsive case, the lowest excitations being again  $\pi$-periodic.

It is proven below
that the $\pi$-periodicity of the lowest-lying excitations is a model independent feature and holds
for all one-dimensional isotropic Heisenberg models having a half-integer value of the spin per unit cell, and 
for all one-dimensional isotropic Hubbard models with an odd number of electrons per unit cell. For Hubbard models at fractional fillings the spectrum of 
lowest excitations is periodic with a period $\pi n$ where $n$ is the number of 
electrons per unit cell. For systems with a nonzero groundstate magnetization
we are able to prove a stronger statement.
We call a 1-D system {\it magnetic in the thermodynamic limit}, 
if there exist such $N_0$, that, for all $N>N_0$, one has 
 $S_N/N \ge  \epsilon>0$, where 
$S_N$ is the spin of the groundstate of a $N$-site ring.\footnote{If the groundstate is nontrivially degenerate, then this condition must be satisfied for 
at least one of groundstates.}. We prove, that if a generalized Heisenberg or 
Hubbard model is magnetic in the thermodynamic limit, then the spectrum
of excitations is gapless at any wave vector $k$, i.e., that, for any $k$, there
exists an excitation of arbitrarily small energy.  
The only restriction on the interaction   
is, that the  
interaction is short-ranged enough, namely, that it falls off more rapidly as $1/r$.
In 1933 Felix Bloch  \cite{bloch} formulated his theorem stating the absence of a stationary current in the 
ground state of a solid with no external field. The physical justification of it is 
straightforward. Suppose the ground state of a solid has got a nonzero current. Let us  make a closed  
electric circuit, consisting
of the solid and some dissipative device (resistance). Then after a relaxation time the current will disappear
and some amount of  energy will be absorbed in the resistance. Since the energy is conserved, the
final zero-current state of the solid will have  energy less than the initial state.
A mathematical proof of the Bloch theorem was given by Bohm in \cite{bohm}.
The Bloch theorem can be also formulated for the spin current, i.e., the current of the
$z$-component of the spin. Both of the versions of the Bloch theorem have been widely used in
solid state physics, for instance  in the theory of superconductivity \cite{abrikosov}. 
It is  proven in this paper, that the spin current (current of the z-component of the spin)
is zero for Heisenberg and Hubbard models in any dimensions.   
To prove our assertions we use the transformation introduced in \cite{lieb}
to prove  the vanishing  of the excitation gap in the thermodynamic limit of half-integer
spin Heisenberg models. This transformation is an analogue
of the transformation  \cite{bohm}  for the case of lattice spin models. 
Consider  first a one-dimensional half-integer spin  Heisenberg model  on a periodic ring, having $2N$ sites.

\begin{equation}
\label{1}
H =  J \sum  {\bf   S}_n \cdot {\bf S}_{n+1} = J  \sum  \left[ S^z_n S^z_{n+1} + \frac{1}{2}\left(  S^{+}_n S^{-}_{n+1} 
+ S^{+}_{n+1} S^{-}_n\right)  \right]
\end{equation}

The sign of $J$ is not specified, therefore both ferromagnetic and antiferromagnetic
cases are considered. The system is translationary invariant and the wave vector $k$ is a good quantum number. For a finite
ring, $k$ takes values $\frac{2\pi}{2N}l$, where $l$ is an integer.  
The lowest energy excitations of the system are defined as excitations minimizing the  energy for a fixed 
value of $k$.  
Below we prove that
the spectrum of the lowest energy excitations $E_{low}\left( k\right) $ is a $\pi$-periodic function of $k$, 
$E_{low}\left( k\right)  = E_{low}\left( k+\pi\right) $.
The system has full rotational symmetry, therefore it is enough to consider the excitations having $S_{\textrm{full}}^z = 0$. 
All other excitations can be produced from $S_{\textrm{full}}^z =0$ by applying operators $S_{\textrm{full}}^+,S_{\textrm{full}}^-$. 
Consider an eigenstate $ \mid \psi_k\rangle$ of $H$, having  wave vector $k$ and  energy $E$.
\begin{equation}
\label{2}
T  \mid \psi_k \rangle  = e^{ik}  \mid \psi_k\rangle,    \; \; \; \; H  \mid \psi_k \rangle  = E  \mid \psi_k\rangle  
\end{equation} 
Here $T$ is the translation operator, $n \rightarrow n+1$.
Let us write the state $ \mid \psi_k\rangle $ as a linear combination of $S_{\textrm{full}}^z=0$ spin configurations
$\psi_k = \sum_\sigma A_\sigma \mid \sigma \rangle$ .

Any $S_{\textrm{full}}^z=0$  configuration
is characterized by  $2 N S$ numbers $x_1 \le x_2 \le x_3 ...$ showing the positions of fictitious particles, each of them increasing  $S^z$ on a given site by one.
The vacuum of the system is assumed to have all spins down. Since the system is periodic, the $x$'s are defined only $mod$ $2N$.
Consider the state 

\begin{equation}
\label{4}
\psi_{k+\pi} = \sum_\sigma e^{\frac{2\pi i}{2N}\left( x_1+x_2+...\right) }  A_\sigma \mid \sigma \rangle . 
\end{equation}
Adding $2N$ to any of the $x$'s does not change  the  $\psi_{k+\pi}$. Therefore 
the mentioned freedom of defining the $x$'s is satisfied.
If  the operator $T$ acts on the state  $\psi_{k+\pi}$, then all $x$'s are incremented by 1. Therefore an
additional phase factor $e^{i\pi 2S}=e^{i \pi}$ is acquired (2S is odd), and the state $\psi_{k+\pi}$ has a wave vector $k+\pi$.
In the same fashion the state
\begin{equation}
\label{5}
\psi_{k-\pi} = \sum_\sigma e^{-\frac{2\pi i}{2N}\left( x_1+x_2+...\right) }  A_\sigma \mid \sigma \rangle 
\end{equation}
has a  wave vector $k-\pi$.
Note that  $k+\pi$ and $k-\pi$ are the same wave vectors since they differ by $2 \pi$.
Let us evaluate the expectation value  $\langle \psi_{k+\pi} \mid H \mid  \psi_{k+\pi} \rangle $.
The Ising part of the Heisenberg Hamiltonian does not flip spins.  In contrast the
exchange part may change the position of one spin-up fictitious particle, which  amounts to a phase factor
$e^{\pm \frac{2\pi i}{2N}}$, depending on a direction of the move.  This explains the result
of a  formal calculation, which gives: 
\begin{eqnarray}
\label{6}
\langle \psi_{k+\pi} \mid H \mid  \psi_{k+\pi} \rangle  = \\ 
\nonumber
 = \sum_n J \left[ 1/2 \left(  e^{\frac{2\pi i}{2N}}  \langle \psi_{k} \mid  S^+_n S^-_{n+1} \mid  \psi_{k} \rangle  +
  e^{-\frac{2\pi i}{2N}}  \langle \psi_{k} \mid  S^+_{n+1} S^-_n  
\mid  \psi_{k} \rangle  \right)  +
  \langle \psi_{k} \mid  S^z_{n+1} S^z_n \mid  \psi_{k} \rangle  \right].
\end{eqnarray}
To understand the physical meaning of the operator $S^+_n S^-_{n+1} - S^+_{n+1} S^-_n$
let us find the time derivative of the $S^z_n$. Due to basic
principles of quantum mechanics one has \cite{landau}
\begin{equation}
\frac{d}{dt} S_n^z = \frac {i}{\hbar} [H,S_n^z] = \frac{i J}{2 \hbar} \left( S^+_{n-1} S^-_n - S^+_n S^-_{n-1}\right)  -
\frac{i J}{2 \hbar} \left( S^+_n S^-_{n+1} - S^+_{n+1} S^-_n\right)  =
{\mathcal J}^z_n - {\mathcal J}^z_{n+1}
\end{equation}
 where
${\mathcal J}^z_n=   \frac{i J}{2 \hbar} \left( S^+_{n-1} S^-_n - S^+_n S^-_{n-1}\right) $. Therefore the operator
${\mathcal J}^z_n=   \frac{i J}{2 \hbar} \left( S^+_{n-1} S^-_n - S^+_n S^-_{n-1}\right) $ has a meaning of the
current  of the $z$ component of the spin through the bond $n-1,n$. The
time derivative of $S^z_n$ is then  a difference of incoming and outcoming currents.
For any eigenstate of the Hamiltonian $H$ all expectation values are stationary, therefore 
$\frac{d}{dt} \langle S_n^z\rangle =\langle {\mathcal J}^z_n\rangle  - \langle {\mathcal J}^z_{n+1}\rangle  =0$.
As a result for any
eigenstate of $H$ the expectation value  $\langle {\mathcal J}^z_n\rangle $  does not depend on $n$.
The spin-current operator $\vec{{\bf {\mathcal J}}}_n$  is a pseudovector with components ${\mathcal J}_n^x,{\mathcal J}_n^y,{\mathcal J}_n^z$. 

For the states having a definite value of $S_{\textrm{full}}^z$ it is 
\begin{equation}
\label{8}
\langle {\mathcal J}_n^x\rangle = \langle {\mathcal J}_n^y\rangle = 0 .  
\end{equation}
This is the selection rule for vector operators discussed in \S 29 of \cite{landau}.
Taking a thermodynamic limit of (\ref{6}) one finds
\begin{eqnarray}
\label{9}
\langle \psi_{k+\pi} \mid H \mid  \psi_{k+\pi} \rangle  - \langle \psi_{k} \mid H \mid  \psi_{k} \rangle  = 
2N \frac{J}{2} (e^{- \frac{2\pi i}{2N}}  \langle \psi_{k} \mid  S^+_{n+1} S^-_n \mid  \psi_{k} \rangle  +
e^{\frac{2\pi i}{2N}}  \langle \psi_{k} \mid  S^+_n S^-_{n+1} \mid  \psi_{k} \rangle - \\   
\nonumber
-  \langle \psi_{k} \mid  S^+_{n+1} S^-_n \mid  \psi_{k} \rangle   
- \langle \psi_{k} \mid  S^+_n S^-_{n+1} \mid  \psi_{k} \rangle ) 
=
i J \pi \left( \langle \psi_{k} \mid  S^+_n S^-_{n+1} \mid  \psi_{k} \rangle  - \langle \psi_{k} \mid  S^+_{n+1} S^-_n \mid  \psi_{k} \rangle \right) .  
\end{eqnarray}
In exactly the same way:
\begin{equation}
\label{10} 
\langle \psi_{k-\pi} \mid H \mid  \psi_{k-\pi} \rangle  - \langle \psi_{k} \mid H \mid  \psi_{k} \rangle  =
- i J \pi \left( \langle \psi_{k} \mid  S^+_n S^-_{n+1} \mid  \psi_{k} \rangle  - \langle \psi_{k} \mid  S^+_{n+1} S^-_n \mid
 \psi_{k} \rangle \right). 
\end{equation}
Note that the expressions (\ref{9}), (\ref{10}) differ by sign. Therefore one of the states (\ref{4}), (\ref{5})
has an expectation value of the energy lower or equal than the state $\mid \psi_k \rangle$.  
In other words, having some excitation at the wave vector $k$ one can always construct
an excitation at the wave vector $k+\pi$ having less or equal energy.     
Let us now  take $\psi_k$ to be a minimum energy excitation for a given value 
of the wave vector $k$. As has been proven there exists an excitation
with a wave vector $k+\pi$ having the energy less or equal to the energy of $\psi_k$. 
Therefore:
$
E_{low}\left( k\right)  \ge E_{low}\left( k+\pi\right) 
$.
But we could have also started from the state $k+\pi$ to show that
$
E_{low}\left( k+\pi\right)  \ge E_{low}\left( k\right) 
$.
Therefore 
$
E_{low} \left( k\right) =E_{low} \left( k+\pi\right) 
$,
so the spectrum of the lowest energy excitations is $\pi$-periodic.
Equal sign in the inequalities above are  realized if and only if  $\langle {\mathcal J}^z_n \rangle  = 0$ 
for any state $\psi_{low}(k)$. 
Combining this fact with (\ref{8}) one has $\vec{{\bf \mathcal J}}_n = 0$, for the lowest excitations
of $H$ having $S_{\textrm{full}}^z = 0$. 
Now consider integer spin models. Then the transformations (\ref{4}), (\ref{5}) produce  states with the same wave vector
as the initial state.
Therefore the $\pi$-periodicity does not work. The statement, that the spin current is equal to zero for the lowest-lying
 $S_{\textrm{full}}^z=0$
states is still valid (otherwise we would find a state with the same $k$ having lower energy). 
For the spin-1/2 AFHM case it is easy to understand, why the spin current is equal to zero for the lowest-lying states
in the thermodynamic limit. As has been already said, the lowest-lying excitations are two-spinon excitations,
so they involve  {\it two} spin-1/2 particles. To produce a nonzero spin current through each bond 
in the thermodynamic limit
one would need to excite a finite number of the particles {\it per unit volume(length)}.

Inclusion of more than nearest-neighbor interactions
or next order terms like $\left( {\bf S}_n \cdot {\bf S}_{n+1} \right)^2$
does not change anything in the proof. Let us, for instance,  add to the
Hamiltonian ({\ref{1})  a 
next-nearest-neighbor 
interaction term $\alpha \left( {\bf S}_n \cdot {\bf S}_{n+2} \right)$.
Here $\alpha$ is a coupling constant. Then the equations (\ref{2})-({\ref{5})
are not changed. The right-hand side of the  equation (\ref{6})
acquires an additional term

\begin{eqnarray}
\label{11}
\alpha \sum_n \left[ 1/2 \left(  e^{ 2 \frac{2\pi i}{2N}}  \langle \psi_{k} \mid  S^+_n S^-_{n+2} \mid  \psi_{k} \rangle  +
  e^{-2 \frac{ 2\pi i}{2N}}  \langle \psi_{k} \mid  S^+_{n+2} S^-_n
\mid  \psi_{k} \rangle  \right)  +
\langle \psi_{k} \mid  S^z_{n+2} S^z_n \mid  \psi_{k} \rangle  \right].
\end{eqnarray}

This term is  obtained by a direct calculation, and its physical explanation
is similar to the explanation of the equation
(\ref{6}). The Ising part of  the next-nearest-neighbor term 
$\alpha \left( {\bf S}_n \cdot {\bf S}_{n+2} \right)$
does not flip spins. In contrast to that, the exchange part may change the position
of one spin-up fictitious particle by {\it two} lattice sites, which amounts to
a phase  $e^{\pm 2 \frac{2\pi i}{2N}}$, depending on a direction of the move.
Taking the thermodynamic limit, one has an additional term
\begin{eqnarray}
2 i  \alpha \pi \left( \langle \psi_{k} \mid  S^+_n S^-_{n+2} \mid  \psi_{k} \rangle  - \langle \psi_{k} \mid  S^+_{n+2} S^-_n \mid  \psi_{k} \rangle \right) .
\end{eqnarray}
on the right-hand side of the equation (\ref{9}).
In the  same fashion the equation (\ref{10}) acquires the term
\begin{eqnarray}
- 2 i  \alpha \pi \left( \langle \psi_{k} \mid  S^+_n S^-_{n+2} \mid 
 \psi_{k} \rangle  - \langle \psi_{k} \mid  S^+_{n+2} S^-_n \mid  
\psi_{k} \rangle \right). 
\end{eqnarray} 
Then arguments, identical to those given after equation ({\ref{10}), lead to a 
$\pi$-periodicity of  lowest-lying excitations.
One can now proceed adding next-next-nearest-neighbor interactions, etc.
The proof goes exactly in the same way, as soon as the interaction range
is {\it finite}. If the interaction has a long range behavior, then one must 
require the long-range part to fall off quickly enough.
Namely, assuming that the interaction falls off  as $1/r^{1+\gamma}$,
$\gamma >0$,  one can
successfully repeat the steps (\ref{2})-(\ref{10}) to prove the $\pi$-periodicity
conjecture in this case. I leave a detailed 
proof for this case to a longer paper \cite{kladko}. 
One can also add nonlinear terms like 
$\left( {\bf S}_n \cdot {\bf S}_{n+1} \right)^m$, and, repeating the same
steps, show by a direct calculation, that the $\pi$-periodicity conjecture
holds for this case too. Let us, for instance,  add to the Hamiltonian  
(\ref{1}) a term $\beta \left( {\bf S}_n \cdot {\bf S}_{n+1} \right)^2$.
We can rewrite this term as 
$ 
\left[ S^z_n S^z_{n+1} + \frac{1}{2}\left(  S^{+}_n S^{-}_{n+1}
+ S^{+}_{n+1} S^{-}_n\right)  \right]^2  
$.
This expression contains terms which do not change $x_i$'s, terms that  change
one of $x_i$'s by 1, and terms, like $(S^{+}_{n+1} S^{-}_n)^2$, 
that change positions of {\it two} $x_i$'s by one. An interested reader
might  check again, that the steps (\ref{2})-(\ref{10}) can be straightforwardly
repeated.

To formulate it shortly, for any one-dimensional half-integer-spin per unit cell isotropic  Heisenberg Hamiltonian 
the $\pi$-periodicity conjecture is  true. The proof of the general conjecture goes exactly
in the same way, as the examples considered above, 
and will be given in the longer paper \cite{kladko}.  
One can also note, that the full spherical symmetry is not necessary.
The only facts really used are, that the Hamiltonian is symmetric with respect to a rotation around the z-axis,
and that the lowest energy excitations can be chosen to have $S_{\textrm{full}}^z = 0$. As long as these conditions are satisfied, 
the spectrum of the lowest energy excitations is $\pi$-periodic.  
  
As have already been stated, the  proof works also for  long-range interactions, if the interaction goes
to zero more rapidly than $1/r$. Haldane and Shastry 
introduced an integrable Heisenberg model with an interaction going as $1/r^2$. The spectrum of  lowest spinon 
excitations of this model is given in \cite{haldane}, and is  $\pi$-periodic, which is an example of the $\pi$-periodicity conjecture, considered here.

Let us note, that in some models not only the lowest excitations might 
be $\pi$-periodic, but the groundstate itself can break
the discrete translational symmetry and dimerize. In this case the size
of the unit cell is doubled, and the Brillouin zone is  halved.
Then {\it all} the excitations of the system will possess a $\pi$-periodicity,
which is in this case related to a symmetry breaking and dimerization. 
An example of such a system is a frustrated 1-D spin-one-half Heisenberg 
chain with $J_2 \gg J_1$ , where $J_1,J_2$ are antiferromagnetic nearest-neighbor  
and next-nearest-neighbor couplings. Extensive studies of this model were
performed in \cite{white}. This model may also be viewed as a zigzag chain, with
an intrachain coupling $J_2$ and an interchain coupling $J_1$. 
The model is exactly solvable for $J_2 = 1/2 J_1$ \cite{ghosh}, the groundstate
is dimerized, and the discrete translational symmetry is broken.
Density Matrix Renormalization Group studies (\cite{white}) show,
that the discrete symmetry breaking appears not only in the exactly solvable 
case,
but for all $J_2 > 0.25 J_1$. There are also helical correlations in the 
model, which could lead to an incommensurate order in higher-dimensions,
but which are destroyed in one dimension, leading to a finite correlation
length and to a gapped state.

Let us now turn to Hubbard models. In this case we claim the following 
$2 k_f$-periodicity conjecture to be true. 
For any isotropic one-dimensional extended Hubbard model the spectrum of 
lowest-lying excitations is $2 k_f$-periodic, where $2  k_f= \pi n$ mod $2 \pi $, and $n$ is 
the number of electrons per unit cell.  
Let us note, that for many simple one-dimensional systems this definition of 
$k_f$ coincides with the Fermi wave vector, when the hubbard $U$ is switched
off, but for more complicated systems (for instance for systems with partially 
filled bands), it is not the case.    
Consider first the simplest case of a one-band Hubbard model
\begin{equation}
\label{14}
H = t \sum_{i,\sigma}  ( c_{i,\sigma}^{\dagger} c_{i+1,\sigma} +h.c.) +
U\sum_i ( n_{i,\uparrow} n_{i,\downarrow})\;\;.
\end{equation}
having $n$ electrons per site, $n$ is a rational number, $n = c_1/c_2$, where $c_1$ and $c_2$ are integers.  
We consider a ring of $2 c_2 N$ sites, eventually taking a limit $N 
\rightarrow \infty$.  The number of  electrons on the ring is equal to 
$2 c_1 N$ and is an even number, 
therefore the full spin of the system  takes integer values. Because
of the full rotational symmetry, it is  enough to consider only
the states $S_{\textrm{full}}^z = 0$. All other excitations can be produced from $S_{\textrm{full}}^z =0$ by applying operators $S_{\textrm{full}}^+,S_{\textrm{full}}^-$.
Analogously to (\ref{2}) we consider an 
eigenstate $ \mid \psi_k\rangle$ of $H$, having a wave vector $k$ and energy $E$.
We write the state $ \mid \psi_k\rangle $ as a linear combination of 
$S_{\textrm{full}}^z=0$
electron
configurations 
$
\psi_k = \sum_\sigma A_\sigma \mid \sigma \rangle .
$ 
Each of the configurations $ \mid \sigma \rangle$ 
is characterized by $c_1 N$ numbers
$x_1 < x_2 < x_3 ...$
, showing the positions of 
$c_1 N$ spin-up electrons, 
and by $c_1 N$ numbers 
$x'_1 < x'_2 < x'_3 ...$, showing the positions of spin-down electrons.
Since the system is periodic, these positions are defined only $mod$ $2N c_2 $.
Now, in analogy to (\ref{4}), consider the state 
$
\psi_{k+ 2 k_f} = \sum_\sigma e^{\frac{2\pi i}{2N c_2}\left( x_1+x_2+...\right) }  A_\sigma \mid \sigma \rangle .
$
Adding $2N c_2$ to any of the $x$'s does not change  the  $\psi_{k+\pi}$. Therefore
the mentioned freedom of defining the $x$'s is satisfied. 
If  the operator $T$ acts on the state  $\psi_{k+\pi}$, then all $x$'s are incremented by 1.
Therefore an additional phase factor $e^{i \pi c_1/c_2}= e^{i  2 k_f}$ is acquired,
and the state $\psi_{k+ 2 k_f}$ has a wave vector $k+ 2 k_f$.
Now we need to construct a state, analogous to the state (\ref{5}). Since 
$ k + 2 k_f$ and $k - 2 k_f$ are in general {\it different} wave vectors,
simply taking the state $k - 2 k_f $ will not work. But we note, that since
$n$ is rational, there always exist an integer number $X$, such that 
$k - X 2 k_f$ and $k+ 2 k_f$ are the same wave vectors.
Let us take $X = 2 c_2 -1$. Then $k - X 2 k_f$ = $k - (2 c_2 -1) \pi c_1/c_2 =
k - 2 \pi c_1 + \pi c_1/c_2 = k + \pi c_1/c_2 = k + 2 k_f$.
Consider now the state 
$
\psi_{k - X 2 k_f} = \sum_\sigma e^{- X \frac{2\pi i}{2N c_2}\left( x_1+x_2+...\right) }  A_\sigma \mid \sigma \rangle .
$
This state has a wave vector $k - X 2 k_f$, which is, as we already said, the 
same as $k + 2 k_f$.   

Let us now evaluate the expectation value $\langle \psi_{k+ 2 k_f} \mid H \mid  \psi_{k+ 2 k_f} \rangle $.
The interaction $U$, and the spin-down hopping do not change the positions 
of spin-up electrons. In contrast the spin-up hopping may change the position of one spin-up electron by one, which  amounts to a phase factor
$e^{\pm \frac{2\pi i}{2 c_2 N}}$, depending on a direction of the move.
 This explains the result
of a  direct calculation, which gives, that in the thermodynamic limit $N 
\rightarrow \infty$ one has 
\begin{eqnarray}
\label{18}
\langle \psi_{k+ 2k_f } \mid H \mid  \psi_{k+ 2 k_f} \rangle  - \langle \psi_{k} \mid H \mid  \psi_{k} \rangle  =
=
i t  \pi \left( \langle \psi_{k} \mid  c^+_{n,\uparrow} c_{n+1,\uparrow} \mid  \psi_{k} \rangle  - \langle \psi_{k} \mid  c^+_{n+1, \uparrow} c_{n,\uparrow} \mid  \psi_{k} \rangle \right) .
\end{eqnarray}
\begin{eqnarray}
\label{19}
\langle \psi_{k - X 2k_f } \mid H \mid  \psi_{k - X  2 k_f} \rangle  - \langle \psi_{k} \mid H \mid  \psi_{k} \rangle  =
=
- X i t  \pi \left( \langle \psi_{k} \mid  c^+_{n,\uparrow} c_{n+1,\uparrow}  
\mid  \psi_{k} \rangle  - \langle \psi_{k} \mid  c^+_{n+1, \uparrow} c_{n,\uparrow} \mid  \psi_{k} \rangle \right) .
\end{eqnarray}

We see, that, since $X$ is by construction a positive integer, if one of the
above expressions is positive, the other is negative, and vice versa.
Therefore one of the above states has the expectation value of energy 
less or equal to the energy of the state $ \mid \psi_k \rangle$.  
Note that both of considered states have the wave vector $k + 2 k_f$.
Now
repeating arguments given after the expression (\ref{10}), we state that
the lowest excitations are periodic functions with a period $2 k_f$.
The $2 k_f$-conjecture is therefore proven.
It is again possible to check, that adding more than nearest-neighbor hoppings
does not change the picture. One can also add arbitrary intersite interaction terms, of the type $V_k n_i n_{i+k}$, since these terms are diagonal in the
considered basis and drop out of calculations. One can also consider 1-D lattices
with more than one site per unit cell (for instance $n$-leg ladders), to show
again, that the spectrum of excitations is $2 k_f$ periodic, where $2 k_f$ we
define as $\pi n$, $n$ being the number of electrons per unit cell. 

So far we used for our derivations the subspace $S_{\textrm{full}}^z = 0$.   
Consider now  a Heisenberg  model, where one of the (possibly many) ground states is magnetic in
the thermodynamic limit,i.e., $\lim_{N \rightarrow \infty} S_{\textrm{full}} / 2N  \ge  M
 >  0$, where $S_{\textrm{full}}$ is the spin of the
ground state. Then for the fixed $N$ the  $S_{\textrm{full}}^z$ of the ground state can be chosen to be from $- 2M N$ to 
$2 M
 N$. Let us choose the ground state to have $S_{\textrm{full}}^z=C$, and write it in a linear combination of $S_{\textrm{full}}^z = C$ configurations.
Let us now consider the state    
$
\mid \psi' \rangle = \sum_\sigma e^{G \frac{2\pi i}{2N}\left( x_1+x_2+...\right) }  A_\sigma \mid \sigma \rangle
$
,
 where $x_1,x_2,...$ are again the positions of the $2 N S + 2 C$ fictitious  particles. $G$ is an integer number.
The wave vector of the state $|\psi'>$ differs from the wave vector of the ground state by $2\pi G (S + C/N)$.
Now choosing different (allowed) values of $C$ and $G$,  
one can show after a careful analysis (which will be given in details in \cite{kladko}), that in the thermodynamic limit there is no gap at any
value of $k$. This result does not depend on whether the spin is half-integer or integer. Therefore if the system
is magnetic in the thermodynamic limit, the spectrum is gapless for any $k$. A similar result can be obtained for one-dimensional 
extended Hubbard models \cite{kladko}.

I thank Prof. Peter Fulde for advice and support and Dr. Sergej Flach, Dr. Stefan Ketteman, Prof. Giniyat Khaliullin,
Dr. Igor Krasovsky, Prof. Valery Pokrovsky,  
Dr. Walter Stephan and Prof. Paul Wiegmann
for important comments.

\end{document}